\documentclass{emulateapj}

\newcommand{\lya}{Ly$\alpha$}
\newcommand{\nv}{N\,{\sc v}}
\newcommand{\oii}{[O\,{\sc ii]} $\lambda$3727}
\newcommand{\hb}{H$\beta$}
\newcommand{\oiii}{[O\,{\sc iii]} $\lambda$5007} 
\newcommand{\ha}{H$\alpha$}

\begin{document}

\title{\boldmath KECK SPECTROSCOPY OF LYMAN-BREAK GALAXIES AND ITS 
IMPLICATIONS FOR THE UV-CONTINUUM AND Ly$\alpha$ LUMINOSITY FUNCTIONS 
AT $z>6$}

\author{Linhua Jiang\altaffilmark{1}, Eiichi Egami\altaffilmark{1}, 
	Nobunari Kashikawa\altaffilmark{2}, Gregory Walth\altaffilmark{1},
	Yuichi Matsuda\altaffilmark{3}, Kazuhiro Shimasaku\altaffilmark{4,5},
	Tohru Nagao\altaffilmark{6}, Kazuaki Ota\altaffilmark{7}, 
	and Masami Ouchi\altaffilmark{8}}
\altaffiltext{1}{Steward Observatory, University of Arizona,
  933 North Cherry Avenue, Tucson, AZ 85721}
\altaffiltext{2}{Optical and Infrared Astronomy Division, National 
	Astronomical Observatory, Mitaka, Tokyo 181-8588, Japan}
\altaffiltext{3}{Department of Physics, Durham University, South Road, 
	Durham DH1 3LE}
\altaffiltext{4}{Department of Astronomy, University of Tokyo, Hongo, 
	Tokyo 113-0033, Japan}
\altaffiltext{5}{Research Center for the Early Universe, University
	of Tokyo, Hongo, Tokyo 113-0033, Japan}
\altaffiltext{6}{Research Center for Space and Cosmic Evolution, Ehime 
	University, Bunkyo-cho, Matsuyama 790-8577, Japan}
\altaffiltext{7}{Department of Astronomy, Kyoto University, 
	Kitashirakawa-Oiwake-cho, Sakyo-ku, Kyoto 606-8502, Japan}
\altaffiltext{8}{Institute for Cosmic Ray Research, University of Tokyo,
   5-1-5 Kashiwa-no-Ha, Kashiwa City, Chiba 77-8582, Japan}

\begin{abstract}

We present Keck spectroscopic observations of $z>6$ Lyman-break galaxy (LBG) 
candidates in the Subaru Deep Field (SDF). The candidates were selected as 
$i'$-dropout objects down to $z'=27$ AB magnitudes from an ultra-deep SDF 
$z'$-band image. With the Keck spectroscopy we identified 19 LBGs with 
prominent \lya\ emission lines at $6\le z\le 6.4$. The median value of the 
\lya\ rest-frame equivalent widths (EWs) is $\sim50$ \AA, with four EWs $>100$ 
\AA. This well-defined spectroscopic sample spans a UV-continuum luminosity 
range of $-21.8\le M_{\rm UV}\le -19.5$ ($0.6\sim5\ L^{\ast}_{\rm UV}$) and 
a \lya\ luminosity range of $\rm (0.3\sim3)\times 10^{43}\ erg\ s^{-1}$ 
($0.3\sim3\ L^{\ast}_{\rm Ly\alpha}$).
We derive the UV and \lya\ luminosity functions (LFs) from our 
sample at $\langle z\rangle \sim6.2$ after we correct for sample
incompleteness. We find that our measurement of the UV LF is consistent with 
the results of previous studies based on photometric LBG samples at $5<z<7$.
Our \lya\ LF is also generally in agreement with the results of \lya-emitter
surveys at $z\sim5.7$ and 6.6.
This study shows that deep spectroscopic observations of LBGs can provide 
unique constraints on both the UV and \lya\ LFs at $z>6$.

\end{abstract}


\section{INTRODUCTION}

The last decade saw great progress in our understanding of the distant 
Universe as a number of objects at $z>6$ were discovered, including galaxies
\citep{iye06,van11}, quasars \citep{fan03,wil10}, and $\gamma$-ray bursts 
\citep{hai06,tan09}. They provide key information to study the formation and 
evolution of the earliest galaxies, supermassive black holes, and massive 
stars when the Universe was less than one billion years old.
The observations of these objects such as the detection of Gunn-Peterson 
troughs in quasar spectra \citep{fan06}, together with the measurements of the 
polarization anisotropies in the cosmic microwave background \citep{kom09},
indicate that at $z>6$ we are approaching the epoch of cosmic reionization, 
during which the intergalactic space became transparent to HI-ionizing UV 
photons.

The first $z>6$ galaxies were discovered to be \lya\ emitters (LAEs) at 
$z\sim6.56$ using the narrow-band technique \citep{hu02,kod03}. This technique 
has been an efficient way to find high-redshift galaxies since the work of 
\citet{cow98} and \citet{rho00}. The number of $z>6$ LAEs has increased to 
almost one hundred with a high success rate of spectroscopic confirmation due 
to the presence of strong \lya\ emission lines 
\citep{tan05,iye06,kas06,hu10,ota10,ouc10,cas11,kas11}. Now \lya\ surveys are 
able to detect LAE candidates at $z=7.7$ \citep{hib10,til10}. However, the 
narrow-band technique has its own limitations. Narrow-band filters are built 
to use dark atmospheric windows with little sky OH emission. Such windows are 
rare at the red end of the optical range (even rarer in the near-IR), and are 
also very narrow ($\sim100$ \AA), resulting in small survey volumes. The 
dropout technique \citep{ste93,gia02} does not have these limitations. It has 
produced a substantial number of Lyman-break galaxy (LBG) candidates at $z>6$ 
\citep{bun04,dic04,yan04,bou08}. 
Most recently with the power of the new $HST$ WFC3/IR camera, LBG candidates 
at $z>7$ (up to $z\sim10$) are being routinely found 
\citep[and references therein]{bun10,fin10,oes10,wilk10,yan10,bou11,lor11}, 
although they are mostly too faint to be spectroscopically confirmed by 
current facilities.

With the large sample of LBG candidates, the galaxy UV luminosity function 
(LF) at $z>6$ has been established. The general result is that the faint-end
slope of the LF at $z\sim6$ is very steep, and the characteristic luminosity
dims significantly from lower redshifts to $z\sim6$ \citep{bou07,mcl09}.
Now one of the main concerns is that there is a lack of well-defined 
spectroscopic LBG samples to cross-check the LF. As mentioned above, 
spectroscopic identifications of $z>6$ LBG candidates are extremely difficult 
unless they are bright and have strong \lya\ emission lines, in which case 
spectroscopy could still be costly \citep{sta10,ono11,van11}. 
Unfortunately most of 
the known LBG candidates at $z>6$ are in the Hubble Ultra Deep Field (HUDF) 
due to the abundance of the high-quality deep data, and thus they are very 
faint. Even the $HST$ WFC3 early release science data \citep{win11} cover only 
40--50 arcmin$^2$ and bright candidates are rare. With a great depth over an 
effective area of $\sim876$ arcmin$^2$, the Subaru Deep Field 
\citep[SDF;][]{kas04} provides a unique field to search for relatively bright 
LBGs.

The SDF project has been very successful in searching for high-redshift 
galaxies. Taking advantage of an 8-m telescope and a prime-focus camera with a 
large field-of-view (FOV, $34'\times27'$), SDF has an impressive depth 
($27.5\sim28.5$ AB mag) in five broad bands $BVRi'z'$ over a survey area of 
one FOV. Especially noteworthy is the deep observations with three narrow-band 
filters, NB816, NB921, and NB973, corresponding to the detection of LAEs at 
$z\sim5.7$, 6.5, and 7, respectively. So far SDF has spectroscopically 
identified $\sim100$ LAEs at $z\sim5.7$ \citep[e.g.][]{shi06} and $z\sim6.5$ 
\citep[e.g.][]{kas06,kas11}, and a few LAEs at $z\sim7$ \citep{iye06,ota10}. 
SDF has also found a sample of bright LBG candidates at $z>6$ \citep{shi05} up 
to $z>7$ \citep{ouc09}. Five strong LAEs from a list of $i'$-dropout objects 
(or LBG candidates) have already been spectroscopically confirmed at $6<z<6.4$
\citep{nag04,nag05,nag07}. Note that the difference between LAEs and LBGs is
somewhat arbitrary and there is no clear separation line between them, so we 
simply call the galaxies found by the narrow-band technique as LAEs and those 
found by the dropout technique as LBGs. A LBG is also a LAE if it is 
identified to have a strong \lya\ emission line. 

In this paper we present our deep spectroscopic observations of $z>6$ LBG 
candidates in SDF using a new, ultra deep $z'$-band image (29 hour 
integration) for target selection. 
This image allows us to select candidates down to $z'=27$ mag, roughly one mag 
deeper than the LBGs found by \citet{nag04,nag05,nag07}. The structure of the 
paper is as follows. Section 2 briefly describes our selection criteria and 
follow-up observations of galaxy candidates. Section 3 presents the results of 
our spectroscopic observations. We derive the UV-continuum and \lya\ LFs in 
Section 4, and summarize the paper in Section 5.
Throughout the paper we use a $\Lambda$-dominated flat cosmology with H$_0=70$ 
km s$^{-1}$ Mpc$^{-1}$, $\Omega_{m}=0.3$, and $\Omega_{\Lambda}=0.7$.
All magnitudes are on an AB system \citep{oke83}.

\section{OBSERVATIONS AND DATA REDUCTION}

\subsection{Selection of Galaxy Candidates at $z>6$}

We selected $z>6$ galaxy candidates using the SDF broad-band images. The SDF 
public data have a depth of $B=28.45$, $V=27.74$, $R=27.80$, $i'=27.43$, 
$\rm NB921=26.54$, and $z'=26.62$ ($3\sigma$ detection for point sources), 
covering an effective area of $\sim876$ arcmin$^2$. \citet{nag04,nag05,nag07} 
have used the public data to find bright LBGs down to $z'\sim26.1$.
Recently the SDF team has obtained a much deeper $z'$-band image with a total
integration time of $\sim29$ hours and a depth of $\sim27.5$ mag
\citep{poz07,ric09}. Our candidate 
selection was based on this deep $z'$-band image together with the public data 
in the other four bluer broad bands.

We used the traditional dropout technique, i.e., our candidates are $i'$-band 
dropout objects. The basic criteria are 
\begin{equation}
	z'<27\ {\rm and}\ i'-z'>1.7.
\end{equation}
The color cut is more stringent than $i'-z'>1.5$ used by 
\citet{nag04,nag05,nag07}. This is to reduce the number of contaminants 
scattered into the selection region due to large $i'$-band photometric errors.
To remove foreground 
contaminants, we required that the candidates are not detected ($<2\sigma$) in 
three broad bands $BVR$. We also rejected possible $z\sim6.56$ LAE candidates 
which are relatively bright in the NB921 band with respect to their $z$-band 
photometry($z'-{\rm NB921}<1$), since these candidates were being targeted in 
another program \citep{kas11}. We obtained 499 $i'$-dropouts in the 
whole SDF field. We then visually inspected each candidate, and removed those 
with any possible detections in any of the $BVR$ bands and those that were 
likely spurious detections (e.g. blended with bright stars). The $i'$-band 
image is deep enough in most cases of our selection. In the extreme case of 
$z'=27$ in which the color cut determines $i'>28.7$,
we visually inspected the candidates and simply required that the candidates 
should not be detected in the $i'$-band (in addition to the $BVR$ bands).
We generated an $i'=28.7$ point source (almost all known $z\ge6$ SDF galaxies 
are point sources in SDF images) and put it in a number of blank regions of 
the $i'$-band image. We cut out these regions and mixed them with other blank
regions that do not have the simulated point source. As a result, more than
95\% of the regions with the $i'=28.7$ source were visually identified, so
our visual inspection is reliable in this case.
Finally, 196 promising candidates survived for our follow-up spectroscopy.

\subsection{Keck/DEIMOS Spectroscopy}

The follow-up spectroscopic observations were carried out with DEIMOS 
\citep{fab03} on the Keck II telescope on 25--28 April 2009. The typical 
seeing was $1\arcsec$. A total of 79 galaxy candidates from the above were 
covered by six masks, but only 73 of them were observed due to slit conflict. 
There were roughly 100 slitlets per mask; most of the slitlets were assigned
to the targets of \citet{kas11} and various secondary targets. We used the 
830 lines mm$^{-1}$ grating with the order blocking filter OG550. The 
wavelength coverage is roughly from 6000 to 10,000 \AA. With a $1\arcsec$ slit 
width, the resolving power was $\sim3600$. The total integration time per mask 
was $\sim3$ hours, broken into individual exposures of 20 or 30 min. We also
observed a spectrophotometric standard star BD+28d4211 in long-slit mode with
the same grating and order blocking filter.
The data were reduced with the DEEP2 DEIMOS data pipeline based on the {\tt 
spec2d} IDL package\footnote{The analysis pipeline used to reduce the DEIMOS 
data was developed at UC Berkeley with support from NSF grant AST-0071048.}.
The DEIMOS flexure compensation system (FCS) failed in the beginning of our
observing run, which caused problems with the data reduction, as flexure could
happen in both spatial and spectral directions. We shifted the spectral images 
along the both directions manually based on sky emission lines and flat-field
images before we fed the data to the pipeline. The pipeline corrected for any
small residual shifts.

We extracted the spectra of our targets and flux-calibrated the spectra using 
alignment stars. There were typically 4--5 bright ($16<R<17$) alignment stars 
per mask used to align masks. They were put in $4\arcsec$ square {\em boxes} 
rather than $1\arcsec$ {\em slits}. The advantage of using alignment stars 
for flux calibration is that alignment stars and the targets were observed 
under exactly the same conditions such as transparency and airmass. For each 
mask, we first measured the spectral response from the spectra of standard 
star BD+28d4211, and extracted the spectra of the alignment stars. 
We then calculated the count-to-flux (erg s$^{-1}$ cm$^{-2}$ \AA$^{-1}$) 
ratios by scaling the spectra of the alignment stars to their broad-band 
photometry ($i'$ or $z'$). The ratios among different alignment stars on a 
mask agree within 0.1 mag. We also incorporated slit loss ($\sim0.24$) into 
the count-to-flux conversion ratios. The slit loss was estimated by assuming
a slit width of $1\arcsec$ and a stable Gaussian PSF of $1\arcsec$ (typical 
seeing). Finally the average of conversion ratios was applied to the spectra 
of other objects in this mask. 
Because the PSF is comparable to the slit width, the slit loss varies with
varying seeing, possible offsets between targets and slits, and even the sizes
of targets. We did not correct for these minor changes.

\section{RESULTS}

Among six mask images, one (the first one during which FCS failed) had more 
than twice lower slit throughput due to the failure of FCS or other unknown 
reasons. We identified one very bright LAE (out of 14 candidates) in this mask
(No. 19 in Table 1 and Figures 1 and 2, see the following paragraphs), but we 
will not include this one in the analysis of spatial density and LF in the 
next section. In the other five masks, we identified 18 galaxies 
(out of 59 candidates) with prominent \lya\ emission lines. 
Our $3\sigma$ detection limit is about
$0.7 \times 10^{-17}$ erg s$^{-1}$ cm$^{-2}$ at $\sim8500$ \AA, and is 
estimated as follows. The detection limit depends on the shape of the \lya\
lines, as narrower lines (for a given flux) are easier to identify.
\citet{kas11} generated two composite \lya\ emission lines for their 
$z\sim5.7$ and 6.5 LAEs and found that the two lines were very similar.
We create a \lya\ emission model image with the shape of the composite 
$z\sim6.5$ LAE profile. We then scale this model image and put a number of
them onto the spectral images. The flux limit is determined by detecting
these simulated \lya\ emission from the spectral images.
Briefly we found one galaxy out of 14 candidates in the mask with very low
throughput and found 18 galaxies out of 59 candidates in the other five masks.
The remaining candidates do not show any continuum
emission nor line emission, so we were not able to identify them.

Figure 1 shows the thumbnail images 
of these galaxies in two broad bands $i'$ and $z'$ and one narrow band NB921. 
They were barely detected in the $i'$-band images and totally invisible in 
the $BVR$ bands. Most of them were also barely detected in the NB921 band.
Figure 2 shows the DEIMOS spectra and the redshifts of the 19 galaxies.
The spectra have been flux calibrated and are placed on an absolute flux scale 
using alignment stars. We can clearly see asymmetry in the emission lines of 
relatively bright galaxies. This is the indicator of the \lya\ emission line 
at high redshift due to strong neutral intergalactic medium (IGM) absorption 
blueward of the line. The non-detection in deep $BVR$ images implies that they
are not likely low-redshift contaminants. In addition, the large wavelength 
coverage rules out the possibility that the detected lines are one of the 
\hb, \oiii, or \ha\ lines. The high resolution of the spectra also ensures 
that they are not \oii\ doublets. In a few cases in Figure 2 there are some 
residual sky lines redward of \lya\ that appear like emission lines. They are
not the AGN feature \nv\ $\lambda$1240, because at $z\sim6$ the \nv\ emission 
line is $\sim170$ \AA\ away.

Table 1 lists the galaxy coordinates, redshifts, and $z'$-band magnitudes,
as well as other properties that will be described below. This galaxy sample 
spans a redshift range $6 \le z \le 6.4$ and a magnitude range of 
$25.1 \le z' \le 27$. Redshift for each galaxy is measured from the \lya\ line 
center using a Gaussian profile to fit the top $\sim50$\% of the line from
the peak (the 
rest-frame \lya\ line center is assumed to be 1216 \AA). Note that the no. 13 
galaxy in our list is no. 2 in the photometric sample of \citet{shi05}.

We measure observed \lya\ line fluxes by integrating the \lya\ spectra over
rest-frame 1215.2 (=1216--0.8) to 1217.2 (=1216+1.2) \AA. We do not correct for
IGM absorption blueward of the line. Table 1 shows the flux measurements and 
the \lya\ luminosities derived from the observed fluxes. Most of the galaxies 
in our sample have the \lya\ fluxes in a range of 
$(0.7\sim2.4) \times 10^{-17}$ erg s$^{-1}$ cm$^{-2}$,
comparable to the \lya\ fluxes in the SDF $z\sim6.5$ LAE sample \citep{kas06}.
The strongest \lya\ emission line 
($5.8 \times 10^{-17}$ erg s$^{-1}$ cm$^{-2}$) in our sample is as bright as 
the strongest LAEs from large LAE surveys of \citet{ouc09}, \citet{hu10}, 
and \citet{kas11}, indicating that it represents the bright end 
of LAEs at $z>6$. Table 1 also includes the star formation rates (SFRs) 
estimated from the \lya\ luminosities by
\begin{equation}
	\rm SFR (Ly\alpha) = 9 \times 10^{-43}\ L(Ly\alpha)\ M_{\sun}\ yr^{-1},
\end{equation}
which is based on the relation between SFR and the \ha\ luminosity
\citep{ken98} and the line emission ratio of \lya\ to \ha\ in Case B 
recombination \citep{ost89}. The derived SFRs are less than 10 
$\rm M_{\sun}\ yr^{-1}$ for all but one galaxy.

We estimate the \lya\ rest-frame equivalent widths (EWs) using the observed 
\lya\ fluxes and $z'$-band photometry. Although the narrow-band NB921 
photometry consists of pure continuum flux for objects at $z<6.5$, our 
galaxies are mostly very faint in this band, preventing us from measuring
reliable photometry. The broad $z'$ band includes emission from both continuum 
and \lya. To decompose the $z'$-band photometry we assume that the UV 
continuum slopes ($f_{\lambda} \propto \lambda^\beta$) are $\beta=-2$ 
\citep{bou09}. We further assume that the continuum blueward of \lya\ is 
entirely absorbed by IGM. This is reasonable as seen from $z\sim6$ quasars 
\citep{fan06}. The only free parameter is then the continuum level when an 
observed spectrum (continuum+\lya) is scaled to match the corresponding 
$z'$-band photometry. The results of the \lya\ EWs are listed in Column 8 of 
Table 1. Column 6 shows $M_{1300}$, the absolute AB magnitude of continuum
at rest-frame 1300 \AA. The EW measurements are rough due to uncertainties 
from the broad-band photometry, \lya\ fluxes, and UV slopes. 
We vary the assumed slope by 0.5 and the typical change on the measured EWs 
is smaller than 15\%. For a few galaxies with strong NB921 detections, we
also derive their EWs based on the continua from their NB921 photometry.
The results mostly agree with the above measurements within 25\%. Nevertheless, 
there is no doubt that most of the galaxies in our sample have large EWs. 
The median value of the EWs is 50 \AA, with 4 galaxies having EWs higher 
than 100 \AA. They are on average smaller than those in the bright sample of 
\citet{nag04,nag05,nag07}, whose \lya\ EWs are in the range of 90--240 \AA.

\section{DISCUSSION}

\subsection{Spectral Properties}

We know little about the \lya\ EW distribution of LBGs at $z\sim6$. Current 
facilities can only identify $z\sim6$ LBGs with prominent \lya\ lines, but it 
is known that most LBGs at low redshift do not have strong \lya\ emission. 
At $z=2\sim3$, the \lya\ EW distribution has been well determined based on a 
spectroscopic sample of more than 1000 LBGs \citep{sha03,red08,red09}. A half 
of the LBGs in this sample have \lya\ absorption lines instead of emission 
lines, and only 17\% have \lya\ EWs greater than 30 \AA\ (with $\sim10$ \% 
having EWs $>50$ \AA), while almost all the LBGs in our sample have EWs $>30$ 
\AA. Therefore, LBGs in our $z>6$ sample have much stronger \lya\ emission on 
average. Statistics of \lya\ EWs for LBGs at $z>5$ has been tentatively 
investigated \citep[e.g.][]{sta10,sta11}. 
For example, \citet{sta11} found that at $-21.75<M_{\rm UV}<-20.25$ the 
fractions of LBGs with EWs $>25$ and $>55$ \AA\ are roughly 20\% and 8\%, 
respectively, and at $-20.25<M_{\rm UV}<-18.75$ the two fractions increase
rapidly to $\sim55$\% and $\sim25$\%. Our sample spans a UV luminosity range 
of $-21.8\le M_{1300}\le -19.5$, and the two fractions in our sample are
25\% (15/59) and 14\% (8/59), broadly consistent with the trend shown in
their results.

The \lya\ strength may also vary with luminosity. Some studies have reported 
the inverse relationship between \lya\ EW and UV luminosity 
\citep[e.g.][]{sha03,and06,red08,sta10}, while others claimed that the 
relation is not obvious or there is no such relation \citep[e.g.][]{nil09}.
Figure 3 shows the \lya\ EW as a function of the continuum luminosity 
$M_{1300}$ for our sample. The dashed line demonstrates the detection limit 
for galaxies at $z=6.2$. The apparent strong correlation between EW and 
$M_{1300}$ in our sample is likely due to the nature of flux-limited surveys.
Nevertheless, our sample has a similar luminosity range as the range of the 
$z\sim3$ LBG sample mentioned above. In the next section we will use the EW 
distribution at $z\sim3$ \citep{red09} to correct for sample incompleteness.

\subsection{Sample Completeness}

The galaxies presented in this paper provide a well-defined flux-limited
galaxy sample down to $z'=27$. The effective area is $\sim340$ arcmin$^2$ 
(five DEIMOS mask coverage) and
the redshift range is $6\le z\le6.4$. In the following subsections we 
will correct for sample incompleteness, calculate the spatial density of the
galaxies, and derive the galaxy LF in this redshift range.

Due to various selection criteria that we applied, our sample is incomplete.
The sample completeness is complicated. Here we correct for incompleteness 
which originates from three major steps, source detection, galaxy 
candidate selection, and spectroscopic identification.
The first major incompleteness comes from the detection of sources in the SDF
image. Because the SDF field is crowded, any high-redshift galaxies behind 
foreground objects are not detected. It is straightforward to calculate this 
incompleteness by measuring the fraction of the area occupied by bright 
objects. There is another source of incompleteness due to the fact that 
fainter objects are more difficult to detect. We put a large number of 
simulated galaxies (point sources) in the SDF $z'$-band image and detect them 
using {\tt SExtractor} \citep{ber96} in the way used for our galaxy candidate 
detection. We then measure the completeness as a 
function of magnitude $z'$ by counting the fraction of detections.
The combined completeness is about 75\% at $z'<25$ and drops to $\sim65$ \%
at $z'=27$.

The second major incompleteness comes from the magnitude limit ($z'<27$) and 
color cut ($i'-z'>1.7$) that we applied to the candidate selection. The last 
major incompleteness is from the fact that our sample is biased towards LBGs 
with strong \lya\ emission, as discussed above. We cannot identify LBGs with 
\lya\ fluxes below our detection limit in the DEIMOS spectra. We use a 
selection function to correct these incompletenesses.

The selection function is defined as the probability that a galaxy with a 
given magnitude, redshift, and intrinsic spectral energy distribution (SED) 
meets the criteria of our candidate selection and \lya\ identification. 
By assuming a distribution for the intrinsic SEDs, we calculate the average 
selection probability as a function of magnitude and redshift. To do this, we 
first produce a large set of galaxy spectra for a given ($M_{1300}, z$), using 
the \citet{bru03} stellar population synthesis model with the Salpeter IMF. 
The magnitude $M_{1300}$ is the absolute AB magnitude at rest-frame 1300 \AA, 
and 1300 \AA\ is chosen to be close to the $z'$-band effective wavelength for 
$z\sim6$ galaxies. The input model parameters will be described in detail in
the next paragraph. We then apply IGM absorption to the model spectra. The 
neutral IGM fraction increases dramatically from $z=5.5$ to $z=6.5$, causing
complete Gunn-Peterson troughs in some $z>6$ quasar spectra \citep{fan06}. 
It is thus critical
to predict $i'-z'$ colors. We calculate IGM absorption in the way used by 
\citet{fan01} and \citet{jia08}. Finally we measure the apparent magnitudes 
from the model spectra with the SDF filter transmission curves. We also 
incorporate photometric errors into each band. It is particularly important
for faint objects, as a non-negligible fraction of real galaxies may have been
scattered out of the selection region due to large photometric errors.
The selection probability for this galaxy ($M_{1300}, z$) is then the fraction
of model galaxies that meet all our criteria.

The intrinsic SEDs of galaxies depend on their physical properties such as 
age, metallicity, and dust extinction. However, we know very little about
physical properties of $z>6$ galaxies. We determine input parameters for the 
above synthesis models from our $HST$ and $Spitzer$ observations of 20 
spectroscopically-confirmed LAEs and LBGs at $5.6<z<7$ in SDF (E. Egami 
et al., in preparation). This sample is bright, and the spectroscopic 
redshifts remove one critical free parameter for SED modeling. 
The $HST$ near-IR data provides rest-frame UV photometry to decipher 
the properties of young stellar populations, while the $Spitzer$ IRAC data 
measure the amplitude of the Balmer break and constrain the properties of 
mature populations. We find that there is a wide variety of SEDs among these 
galaxies, such as mature galaxies with ages $>100$ Myr and young galaxies with 
ages $\sim1$ Myr. The dust extinction is low to moderate, consistent with the
general trend that higher-redshift galaxies have bluer UV continuum slope and
lower dust extinction \citep{bou09}. Based on these results, our model 
parameters are set as follows.
At a given redshift (from 5.5 to 7.2), we choose to use a mixed grid of six 
ages [1, 2, 5, 10, 40, 100] Myr and three dust extinction values E(B--V) = 
[0.02, 0.1, 0.3]. The metallicity for the models with ages of 1 and 2 Myr is 
0.005 $Z_{\sun}$, for the models of with ages of 5 and 10 Myr is 0.02 
$Z_{\sun}$, and for the models of with ages of 40 and 100 Myr is 0.2 
$Z_{\sun}$. The selection function is not sensitive to these 
physical parameters such as age, metallicity, and dust extinction.
After we generate the galaxy continuum spectra using the \citet{bru03} model,
we add \lya\ emission lines to the spectra. The \lya\ emission in our galaxies 
is strong, and has significant contribution ($\sim0.4$ mag) to the $z'$-band
photometry. As we discussed in Section 3, we know very little about the 
statistics of \lya\ strengths at $z\sim6$. Therefore we assume that the 
rest-frame EWs of \lya\ at $z\sim6$ have a distribution similar to that of 
$z\sim3$ \citep{red09}.

Figure 4 shows the selection function as a function of $M_{1300}$ and $z$. 
The contours in the figure are selection probabilities from 5\% to 35\% with 
an interval of 5\%. The sharp decrease of the probability at $z\sim6$ is due 
to the color cut of $i'-z'>1.7$. The solid circles are the locations of the 19 
$z>6$ galaxies in our sample. Figure 4 does not include two small constant 
incompletenesses. One is from slit conflict during DEIMOS slit assignment; 
$\sim8$\% candidates were not allocated slits. The other one is due to
the existence of strong sky OH lines; emission lines that happen to be on 
these OH lines are much more difficult to identify. Although in principle this 
incompleteness is also a function of magnitude and redshift, we find that 5\%
is a good approximation for our sample in the range of 8500--9000 \AA\ 
\citep[e.g.][]{sta10}. Figure 4 does not take into account the rejection of 
$z\sim6.5$ LAEs either, otherwise there will be a `dip' at $z\sim6.5$ in the 
selection function of Figure 4.

\subsection{UV Luminosity Function}

We derive the volume density of the galaxies at $z>6$ using the traditional 
$1/V_{a}$ method. The available volume for a galaxy with absolute magnitude 
$M_{1300}$ and redshift $z$ in a magnitude bin $\Delta M$ and a redshift bin 
$\Delta z$ is
\begin{equation}
	V_{a} = \int_{\Delta M}\int_{\Delta z}p(M_{1300},z) \frac{dV}{dz} dz\,dM,
\end{equation}
where $p(M_{1300},z)$ is the probability function used to correct for all the
sample incompletenesses described above. Then the spatial density and its
statistical uncertainty can be written as \citep{pag00}
\begin{equation}
	\rho = \frac{N_{gal}} {V_{a}}, \ \
	\sigma(\rho) = \frac{N_{gal}^{1/2}} {V_{a}},
\end{equation}
where $N_{gal}$ is the number of galaxies in the bin ($\Delta M, \Delta z$).
The magnitude and redshift distributions of the galaxies in our sample are
shown in Figure 4. We measure galaxy densities in one redshift bin 
$5.8<z<6.5$ and three magnitude bins [--19.4, --20.4], [--20.4, --21.4], and 
[--21.4, --23]. The lower limits of redshift and magnitude are chosen to be 
our detection limit (probability in Figure 4 below 5\%). The magnitude upper 
limit --23 of the last bin is an assumption, comparable to the brightest LBG
candidates in the largest LBG sample of \citet{mcl09}. We choose 6.5 as our 
upper limit of redshift because we were not able to identify LBG candidates at 
$z>6.6$ (see subsection 4.5 for discuss) and the LAEs at $z\sim6.56$ are being 
targeting in another program (see Section 2).

Figure 5 shows our measurements of the spatial densities (the filled 
circles with error bars) in the three magnitude bins at 
$\langle z\rangle \sim6.2$. It also compares with the UV LFs at $z=5$, 5.9,
and 6.8 from other studies \citep{bou07,mcl09,ouc09,bou11}. \citet{bou07} 
obtained a large sample of LBG candidates from the HUDF and other $HST$ 
datasets. They derived the UV LFs at $z=5$ and 5.9 (the blue and green 
dashed lines in Figure 5) down to a depth of $M_{UV}\sim-17$ and $-18$, 
respectively, and found very steep faint-end slopes ($\alpha\sim-1.7$) at 
these redshifts. \citet{mcl09} extended the $HST$ sample by including brighter 
LBG candidates selected from the UKIDSS Ultra Deep Survey (UDS). They found 
very similar LFs at $z\sim5$ and 6 (the blue and green dotted lines in Figure 
4). The $z=6.8$ LFs in Figure 5 were measured with the data from the new 
$HST$ WFC3 IR camera \citep{bou11} and with the deep SDF $Y$-band imaging data 
\citep{ouc09}.
It is clear that our spectroscopic result is in good agreement with the trend
of the LFs from $z=5.9$ to 6.8 from the previous studies.
Note that k-correction within a small UV wavelength range ($1300\sim1600$ \AA)
is negligible because of the blue UV slopes.

We parametrize the galaxy UV LF at $z\sim6$ from our spectroscopic sample
using a Schechter function,
\begin{equation}
	\phi(M) = 0.4\, {\rm ln}10\, \phi^{\ast} 10^{-0.4(M-M^{\ast})(\alpha+1)}\, 
	{\rm exp}(-10^{-0.4(M-M^{\ast})}),
\end{equation}
where $\phi^{\ast}$ is the normalization, $M^{\ast}$ is the characteristic
luminosity, and $\alpha$ is the faint-end slope. Because our sample is not 
deep enough to well determine $\alpha$, and the measurements of $\alpha$ at 
$5\le z\le6$ from previous studies are quite robust and consistent, we fix 
$\alpha$ to the value of --1.74 given by \citet{bou07}. We then fit our 
data to the above function. The best fits are 
$\rm \phi^{\ast} = (1.7\pm1.2)\times 10^{-3}\ Mpc^{-3}$ and 
$M^{\ast} = -19.97\pm0.32$, consistent with the result of \citet{bou07} 
($\rm \phi^{\ast} = 1.4\times10^{-3}\ Mpc^{-3}$ and $M^{\ast} = -20.24$) and 
the result of \citet{mcl09} ($\rm \phi^{\ast} = 1.8\times10^{-3}\ Mpc^{-3}$ 
and $M^{\ast} = -20.04$) at $z\sim6$.
The best-fitting $M^{\ast} = -19.97$ indicates that our sample spans a large
luminosity range $0.6\, L^{\ast} \sim 5\, L^{\ast}$ across the 
characteristic luminosity $L^{\ast}$ at $z\sim6.2$. The result is shown in 
Figure 5.

Cosmic variance has a minor effect on our results. The six masks covered
different parts of SDF, and we did not exclusively target denser regions of 
the LBGs candidates. We covered 79 candidates in 340 arcmin$^2$, while we
selected 196 promising candidates in the whole SDF field ($\sim 876$ 
arcmin$^2$). The candidate surface density in the six masks ($0.23=79/340$) is 
close to the average density of the field ($0.22=196/876$). Therefore our 
sample is a representative sample of SDF. We calculate the uncertainty 
originating from
cosmic variance for all the SDF candidates using the \citet{tre08} calculator.
Within a redshift range of $5.8<z<6.5$ the uncertainty is only $\sim20$ \%,
much smaller than statistical errors in $\phi^{\ast}$ and $M^{\ast}$.

Our galaxy sample is a well-defined spectroscopic sample of LBGs at $z>6$. 
The comparison in Figure 5 shows that our measurement of the UV LF 
at $z\sim6$ are consistent with the results from deep $HST$ and UKIDSS 
observations. However, our LF largely depends on the selection function 
described in Figure 4. Our tests show that the selection function is not 
sensitive to the physical parameters such as age, metallicity, and dust 
extinction, but is sensitive to the distribution of \lya\ EWs.
Because the fraction of strong LAEs among LBGs is small, we have applied
a large correction in the selection function. Figure 5 may indicate that 
the distribution of \lya\ strength at $z\sim6$ is similar to that at $z\sim3$. 
Otherwise our LF will be quite diffirent from the previous measurments of the 
$z\sim6$ LFs. Alternatively, the contamination rates in previous LBG samples 
should be small. For example, the contamination rate in the \citet{bou07} 
sample was claimed to be as small as $\sim3$ \%, and in the \citet{bou11} 
sample was found to be less than 14 \%. If their contamination rates are 
higher, the number densities derived from their samples will be lower, in 
contrary to Figure 5.

\subsection{\lya\ Luminosity Function}

We also derive the \lya\ LF using the way very similar to what we do for the 
UV LF. The \lya\ luminosites $L$(\lya) are listed in Table 1. We first 
calculate the selection function to correct for incompleteness for our 
galaxies as a sample of LAEs. Simulated galaxies are generated using the 
\citet{bru03} synthesis model as discrbied in Section 4.2. The only difference 
is the assumption of the \lya\ EW distribution. The statistics of \lya\ 
strengths in LAEs at $z\sim6$ has been determined in several studies. We use 
the EW probability distribution estimated from a sample of photometrically 
selected LAEs of \citet{ouc08}.
Figure 6 shows the LAE selection function as a function of $L$(\lya) and $z$.
The contours in the figure are selection probabilities from 10\% to 70\% with 
an interval of 10\%. The selection function is generally similar to the one
for LBGs in Figure 4, except that the selection probabilities for LAEs are 
more than three times higher than those for LBGs. This is because for LAEs we
do not need to apply a large correction due to the small fraction LAEs among 
LBGs. 

The volume densities of the LAEs are computed in four luminosity (log$\,L$) 
bins [42.35, 42.60], [42.60, 42.85], [42.85, 43.10], and [43.10, 43.50].
The lower limit 42.35 of the first bin is chosen to be our detection limit 
(probability in Figure 6 below 5\%). The brightest bin contains only one
galaxy and the bin size is arbitrary, so we do not include this bin in the
following analysis. We estimate the \lya\ LF using a Schechter function,
\begin{equation}
   \phi({\rm log}\, L) = {\rm ln}10\, \phi^{\ast} (L/L^{\ast})^{\alpha+1}\, 
   {\rm exp}(-L/L^{\ast}).
\end{equation}
Most galaxies in our sample cover the faint end of the \lya\ LF, so we are not 
able to well determine $\,L^{\ast}$. Therefore we fix log$\,L^{\ast}$ to be 
the value of 43.0 \citep[e.g.][]{kas06,hu10,cas11,kas11} when we fit our data 
to the above function. The best fits are
$\rm \phi^{\ast} = (8.2\pm3.3)\times 10^{-5}\ Mpc^{-3}$ and
$\alpha = -1.67\pm0.54$. Figure 7 shows our measurements of the densities 
(filled circles with error bars) and the best model fit (solid curve). As we 
mentioned, the selection function is sensitive to the \lya\ EW distribution, 
especially in the faint end of the sample. The filled triangles in Figure 7 
represent the volume densities if the assumed EW distribution is increased by 
30 \AA\ (to roughly match the EW distribution for the spectroscopic sample 
shown in \citet{ouc08}). The density in the faintest bin is increased by a 
factor of two. In Figure 7 we also show the comparison with the \lya\ LF
measurements at $z\sim6.5$ from \citet{hu10}, \citet{kas11}, and \citet{ouc10}.
The results of \citet{kas11} and \citet{ouc10} agree with each other in a
wide luminosity range covered. The result of \citet{hu10} also agrees with 
them in the bright end of the LF, but show a significant discrepancy in the 
faint end, where the density of \citet{hu10} is about three times lower. 
The reason is still unclear. Our LF in the faint end is slightly higher than 
that of \citet{hu10}, but is lower than the \citet{kas11} and \citet{ouc10}
LFs. Given the large uncertainties
our result is generally in agreement with these previous LAE surveys.

\subsection{Did We Miss $z>6.6$ LBGs?}

The most distant galaxy in our sample is at $z\sim6.4$, and we did not find 
any galaxies at $z>6.6$. The reason is complicated. The deep $z'$-band image 
used in this paper has securely detected one of the most distant LAEs known at 
$z=6.96$ \citep{iye06}. We successfully recovered this object during our 
candidate selection, and we also have six good $z>6.6$ candidates that have 
been observed. If they were galaxies like the $z=6.96$ LAE, we should have 
identified them in the DEIMOS mask images. 
This is illustrated in Figures 4 and 6. The detection probability of the 
$z=6.96$ LAE is higher than those for a half of the LBGs or LAEs in our 
sample. However, the throughput of DEIMOS drops steadily from 9000 \AA\ 
towards higher wavelength, and in this range the sky background is brighter 
and the OH lines are denser. These issues result in significantly lower 
signal-to-noise ratios in the DEIMOS images, but they were not considered in 
Figures 4 and 6. Therefore it is likely that our spectra at $>9000$ \AA\ are 
not deep enough to identify \lya\ emisson lines. On the other hand, we cannot 
rule out the possibility that there are no $z>6.6$ LBGs with strong \lya\ 
emission in the covered region. Due to the increasing neutral fraction of IGM
at $z>6$, the density of LAEs at the bright end may decline rapidly towards 
higher redshifts \citep[e.g.][]{kas11,pen11,sch11}.

\section{SUMMARY}

In this paper we have reported the discovery of 19 LBGs from our deep 
spectroscopic observations of a sample of $z>6$ LBG candidates in SDF. 
The candidates were selected using the traditional dropout technique from 
an ultra-deep $z'$-band image. This image, with a total integration time of
29 hours, enables us to select galaxies down to $z'=27$ mag over a wide field. 
The follow-up spectroscopy was made with Keck DEIMOS. The 19 LBGs span a 
redshift range of $6 \le z \le 6.4$ and a magnitude range of 
$25.1 \le z' \le 27$. 
They have moderate \lya\ emission line strengths compared to those in LAEs at 
similar redshifts. The median value of rest-frame \lya\ EWs is $\sim50$ \AA, 
and four LBGs have EWs $>100$ \AA. 

This well-defined spectroscopic LBG sample spans a UV luminosity range 
of $0.6\sim5\ L^{\ast}_{\rm UV}$ across the UV characteristic luminosity 
$L^{\ast}_{\rm UV}$ and a \lya\ luminosity range of 
$0.3\sim3\ L^{\ast}_{\rm Ly\alpha}$ across the \lya\ characteristic 
luminosity $L^{\ast}_{\rm Ly\alpha}$. 
It thus provides unique constraints on both the UV and \lya\ LFs at $z>6$. We 
correct for sample incompleteness from source 
detection, galaxy candidate selection, and spectroscopic identification. 
In particular, for the UV LF we assume that the distribution of \lya\ EWs at 
$z\sim6$ is the same as that at $z\sim3$. We then calculate the volume
density of the galaxies and estimate the LFs at $\langle z\rangle \sim6.2$ 
using a Schechter function. We find that our measurement of the UV LF is 
consistent with the results of previous studies based on photometric LBG 
samples at $5<z<7$, including samples from HDF and UKIDSS UDF. 
Our \lya\ LF is also in agreement with the results of \lya-emitter
surveys at $z\sim5.7$ and 6.6.

\acknowledgments

We acknowledge the funding support from NASA through awards issued by STScI
(HST PID: 11149) and JPL/Caltech (Spitzer PID: 40026). We thank Xiaohui Fan 
for providing the simulation of IGM absorption.
The imaging data presented herein were collected at Subaru Telescope, which is 
operated by the National Astronomical Observatory of Japan. The spectroscopic
data were obtained at the W.M. Keck Observatory, which is operated as a 
scientific partnership among the California Institute of Technology, the 
University of California, and the National Aeronautics and Space 
Administration. The Observatory was made possible by the generous financial 
support of the W.M. Keck Foundation.
The authors wish to recognize and acknowledge the very significant cultural 
role and reverence that the summit of Mauna Kea has always had within the 
indigenous Hawaiian community. 
Keck telescope time was granted by NOAO, through the Telescope System 
Instrumentation Program (TSIP). TSIP is funded by NSF.

{\it Facilities:} 
\facility{Keck (DEIMOS)}.

\clearpage
\begin{deluxetable}{cccccccccc}\footnotesize
\tablecaption{Properties of the 19 LBGs}
\tablewidth{0pt}
\tablehead{\colhead{} & \colhead{RA} & \colhead{Dec} & \colhead{} &
	\colhead{$z'$} &  \colhead{$M_{1300}$} & \colhead{$f$(\lya)} &
	\colhead{EW(\lya)} & \colhead{$L$(\lya)} & \colhead{SFR(\lya)} \\
	\colhead{No.} & \colhead{(2000)} & \colhead{(2000)} & \colhead{Redshift} &
   \colhead{(mag)} &  \colhead{(mag)} & 
	\colhead{(10$^{-17}$ erg s$^{-1}$ cm$^{-2}$)} &
   \colhead{(\AA)} & \colhead{(10$^{42}$ erg s$^{-1}$)} & 
	\colhead{($M_{\sun}$ yr$^{-1}$)} }
\startdata
 1& 13:25:18.142& 27:32:32.449& 6.240& 26.99& --19.80& 0.8&  47&  3.5&  3.2\\
 2& 13:24:55.589& 27:39:20.772& 6.125& 26.69& --19.63& 1.4&  90&  5.7&  5.2\\
 3& 13:24:41.333& 27:43:16.601& 6.394& 26.47& --20.69& 1.3&  35&  5.9&  5.4\\
 4& 13:24:36.893& 27:43:41.520& 6.343& 26.38& --20.74& 1.2&  29&  5.2&  4.7\\
 5& 13:25:11.086& 27:37:47.957& 6.146& 26.29& --20.51& 1.0&  23&  3.4&  3.1\\
 6& 13:24:26.030& 27:16:02.993& 5.992& 26.88& --19.54& 1.2&  81&  4.7&  4.3\\
 7& 13:24:20.626& 27:16:40.408& 6.267& 26.91& --19.90& 1.0&  50&  4.1&  3.7\\
 8& 13:24:05.894& 27:18:37.703& 6.047& 26.82& --19.74& 0.7&  41&  2.8&  2.6\\
 9& 13:24:10.769& 27:19:03.918& 6.038& 26.61& --19.57& 1.7& 120&  7.2&  6.5\\
10& 13:24:34.824& 27:14:18.985& 6.233& 26.98& --19.67& 1.1&  63&  4.1&  3.8\\
11& 13:25:21.053& 27:15:24.023& 6.076& 26.34& --19.95& 1.9&  92&  7.8&  7.1\\
12& 13:25:22.190& 27:21:41.069& 6.214& 26.56& --19.51& 2.4& 170&  9.6&  8.7\\
13& 13:25:27.804& 27:28:58.742& 6.226& 25.31& --21.77& 1.1&  10&  4.6&  4.2\\
14& 13:23:38.616& 27:26:15.623& 6.051& 26.73& --19.83& 0.7&  41&  3.1&  2.8\\
15& 13:23:45.758& 27:32:51.342& 6.313& 25.59& --21.66& 0.9&  10&  4.2&  3.8\\
16& 13:23:42.137& 27:33:33.905& 6.038& 26.05& --20.58& 1.1&  30&  4.5&  4.1\\
17& 13:23:50.484& 27:34:15.078& 6.302& 26.72& --19.80& 1.6& 101&  7.5&  6.8\\
18& 13:24:26.107& 27:18:40.450& 6.131& 26.59& --19.52& 2.0& 136&  7.7&  7.0\\
19& 13:25:21.612& 27:42:28.915& 6.164& 25.10& --21.25& 5.8&  95& 26.5& 24.1\\
\enddata
\tablecomments{The magnitude $M_{1300}$ is the absolute AB magnitude of 
continuum at rest-frame 1300 \AA. EW(\lya) is the \lya\ rest-frame equivalent 
width. The \lya\ properties are not corrected for IGM absorption.}
\end{deluxetable}

\clearpage
\begin{figure}
\epsscale{0.9}
\plotone{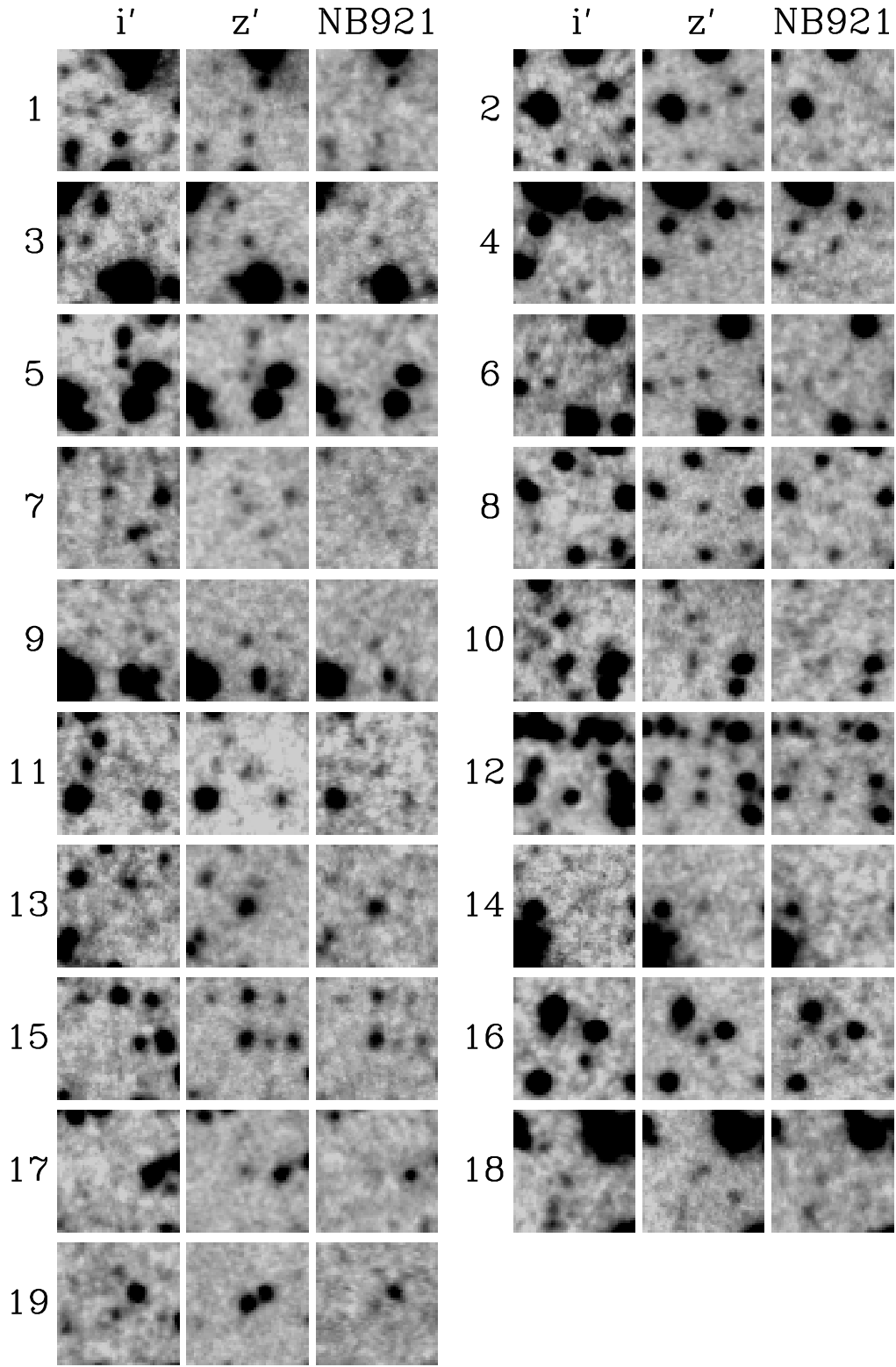}
\caption{Thumbnail images of the 19 LBGs in two broad bands $i'$ and $z'$ and 
one narrow band NB921. The image size is $10\arcsec \times 10\arcsec$. North
is up and east to the left. These galaxies were barely detected in the 
$i'$-band images and totally disappeared in the $BVR$ bands. Most of them were 
also barely detected in the NB921 band.}
\end{figure}

\clearpage
\begin{figure}
\epsscale{0.9}
\plotone{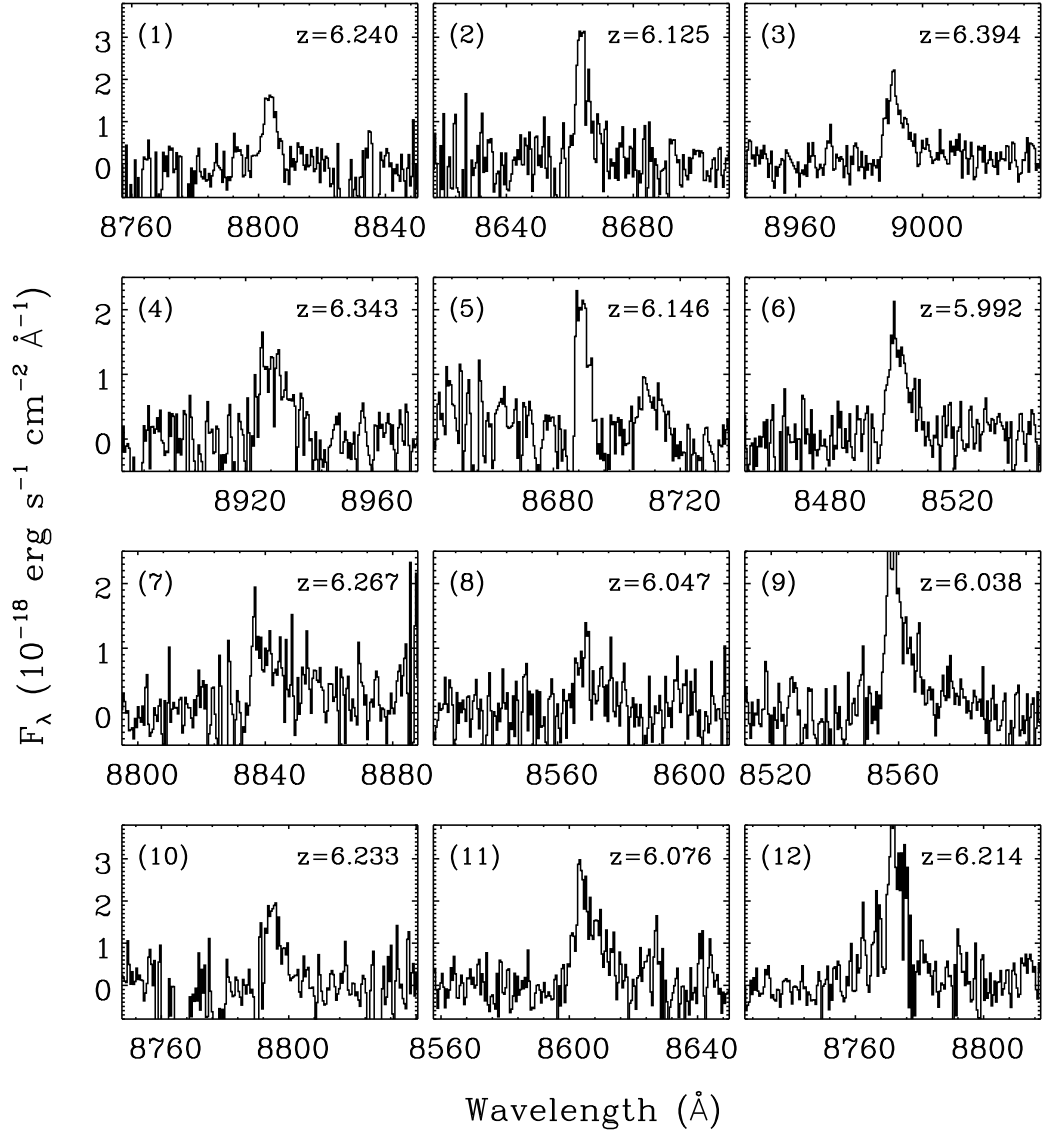}
\caption{Spectra of the 19 LBGs. The \lya\ emission line is shown in the 
center of each panel. The spectra have been flux calibrated and are placed on 
the absolute flux scale.}
\end{figure}

\clearpage
\addtocounter{figure}{-1}
\begin{figure}
\epsscale{0.9}
\plotone{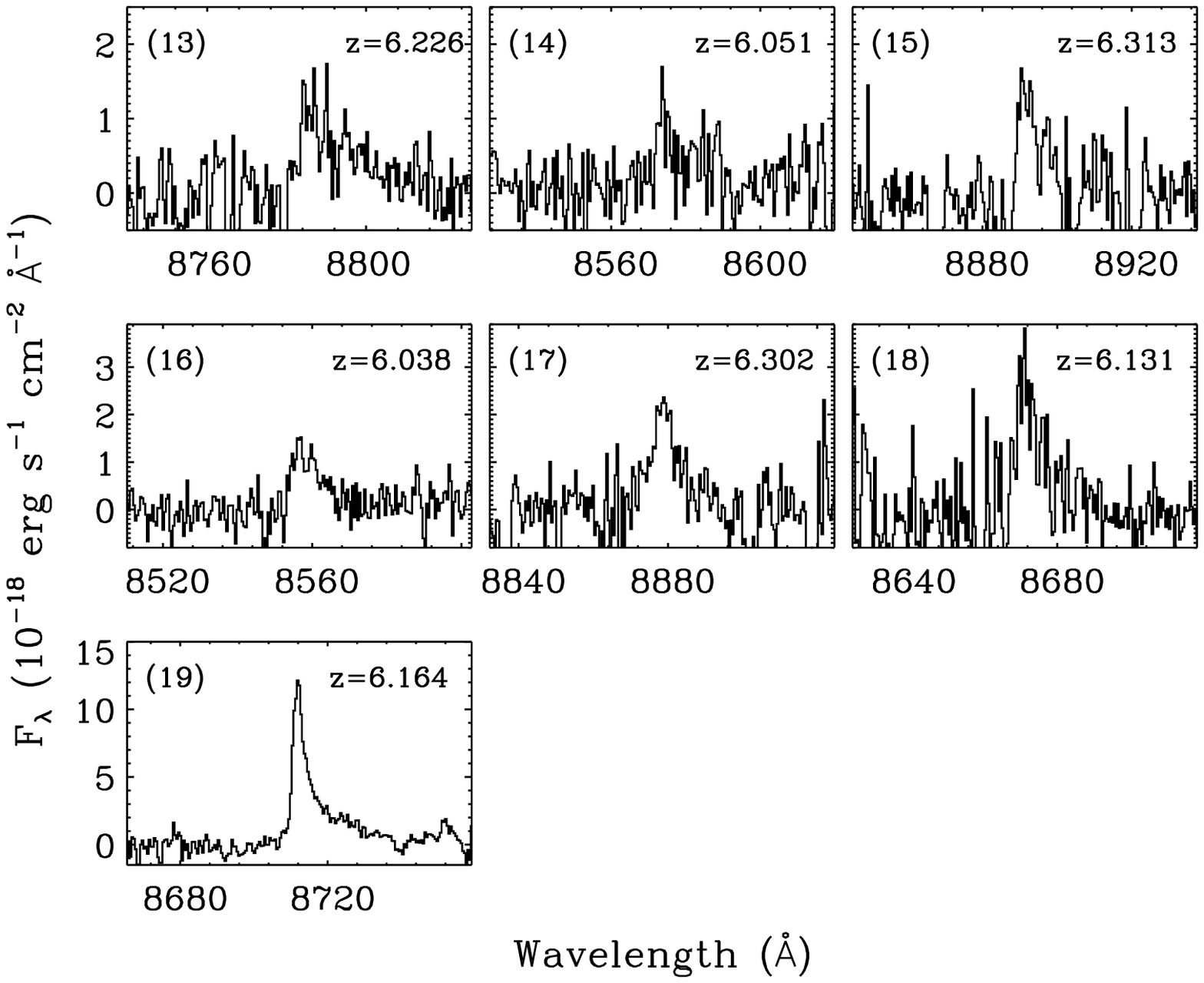}
\caption{Continued.}
\end{figure}

\begin{figure}
\epsscale{0.7}
\plotone{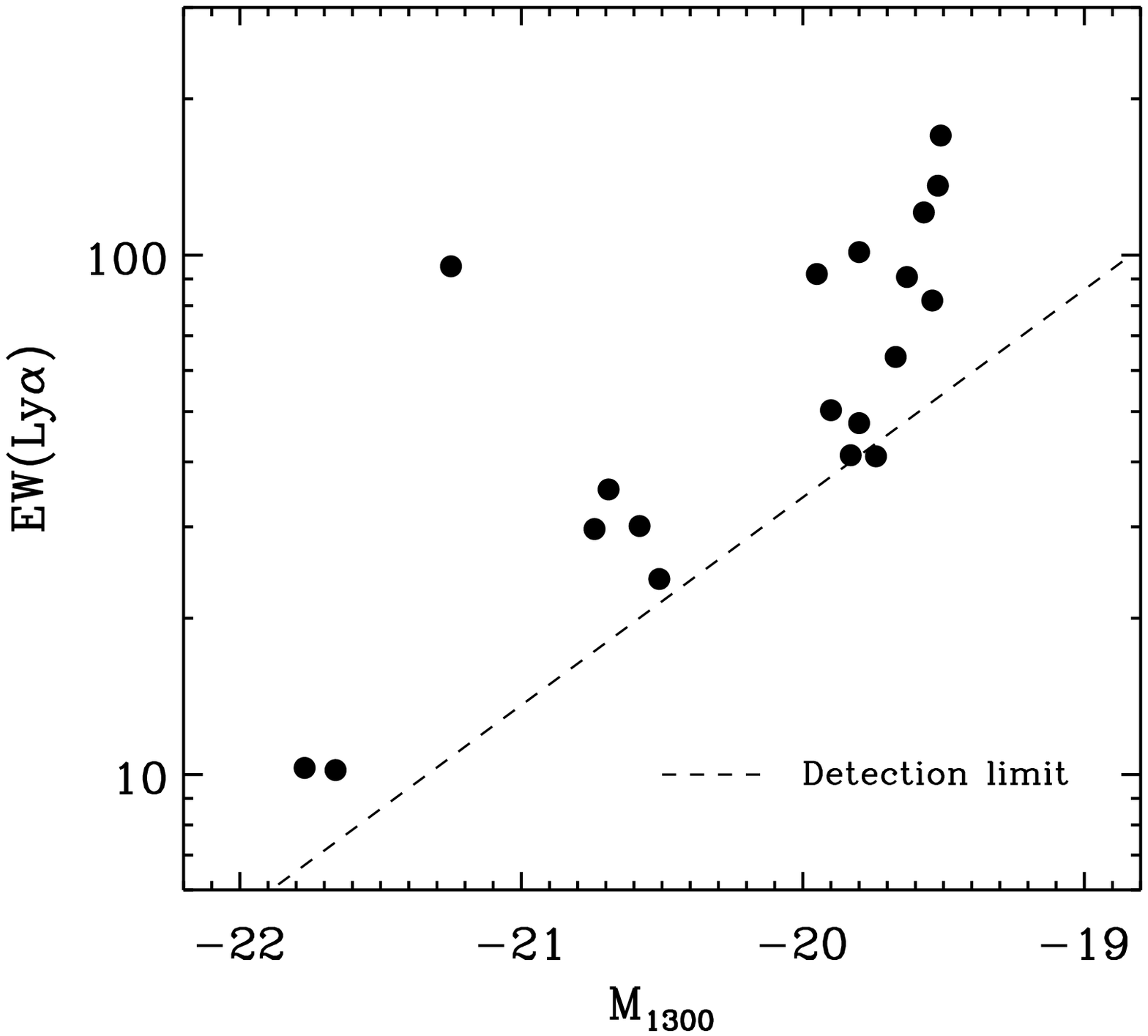}
\caption{\lya\ EW as a function of the continuum luminosity $M_{1300}$ for our 
sample. The filled circles represent the 19 galaxies at $6<z<6.4$. The dashed 
line demonstrates our detection limit for galaxies at $z=6.2$. 
The strong relation between EW and $M_{1300}$ in our sample is very likely due 
to the nature of flux-limited surveys.}
\end{figure}

\clearpage
\begin{figure}
\epsscale{0.7}
\plotone{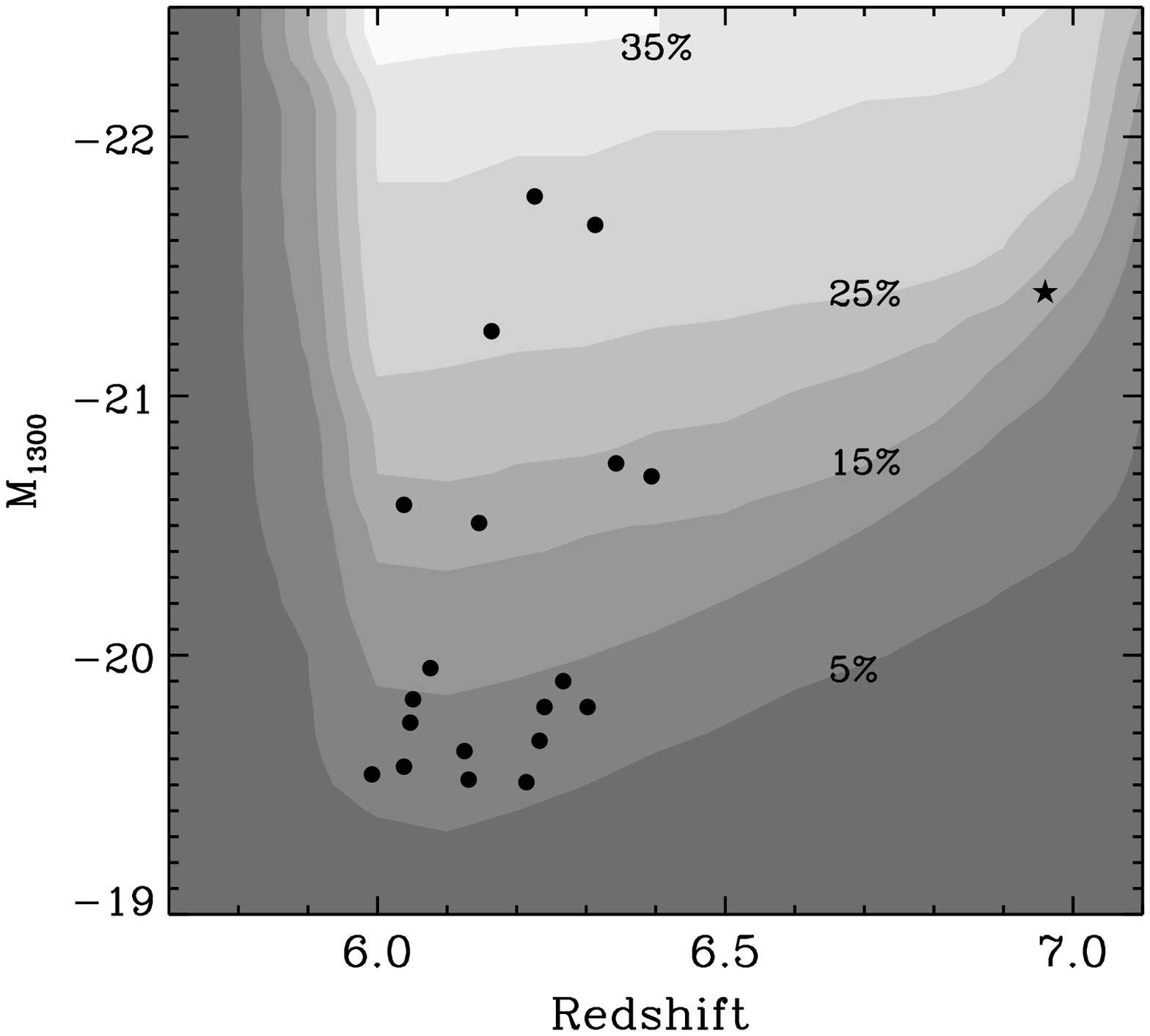}
\caption{LBG selection function as a function of $M_{1300}$ and $z$.
The contours are selection probabilities from 5\% to 35\% with an 
interval of 5\%. The filled circles are the locations of the 19 $z>6$ 
galaxies in our sample. The filled star is the $z=6.96$ LAE \citep{iye06}.}
\end{figure}

\begin{figure}
\epsscale{0.7}
\plotone{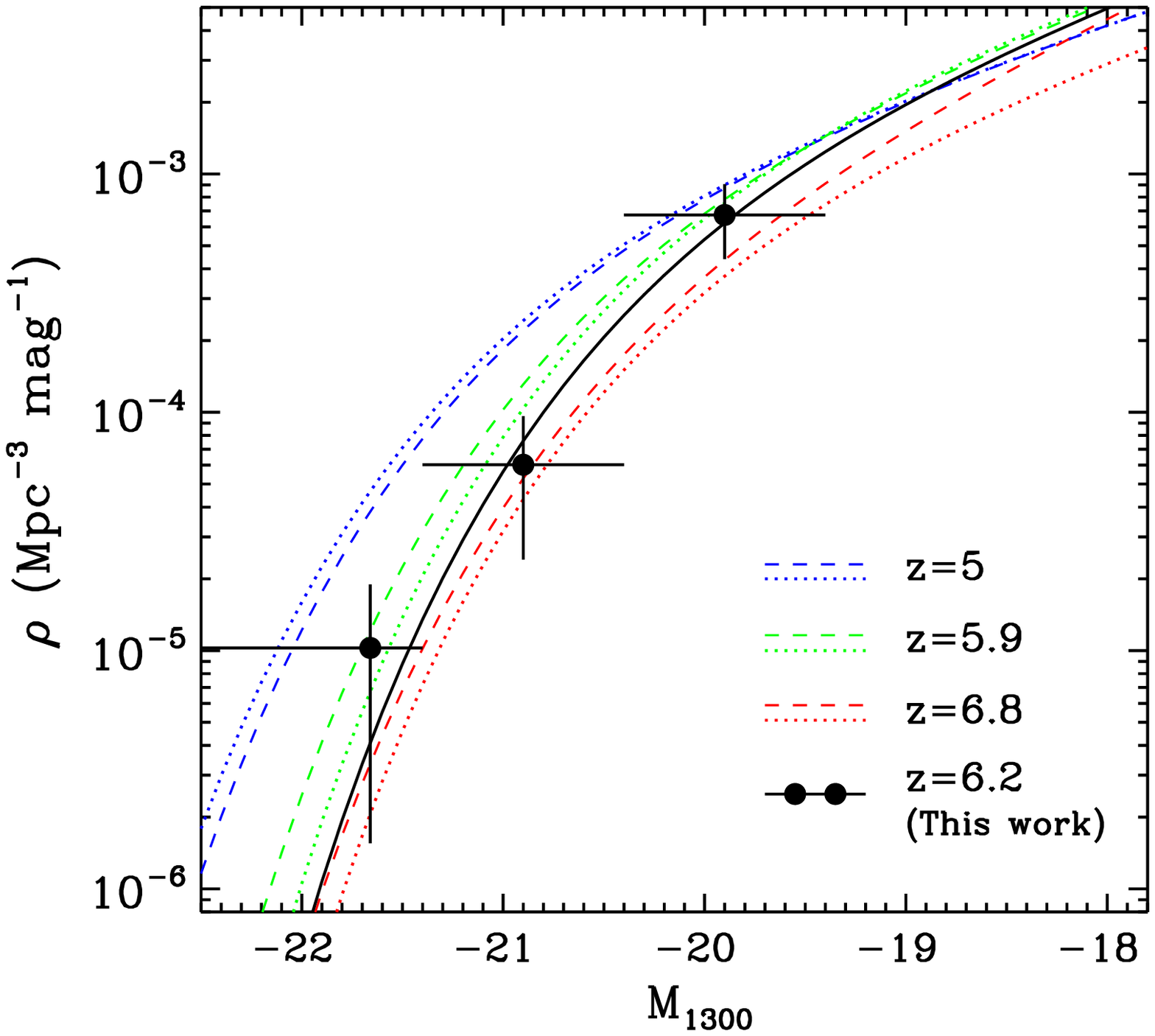}
\caption{UV LF of LBGs at high redshift. The three filled circles are the 
spatial densities measured from our sample, and the solid curve is the best 
fit to a Schechter function. As comparison, color-coded dashed and dotted 
lines represent UV LFs from previous studies based on photometric LBG samples.
The blue and green dashed lines: \citet{bou07}. 
The blue and green dotted lines: \citet{mcl09}.
The red dashed line: \citet{bou11}.
The red dotted line: \citet{ouc09}.}
\end{figure}

\clearpage
\begin{figure}
\epsscale{0.7}
\plotone{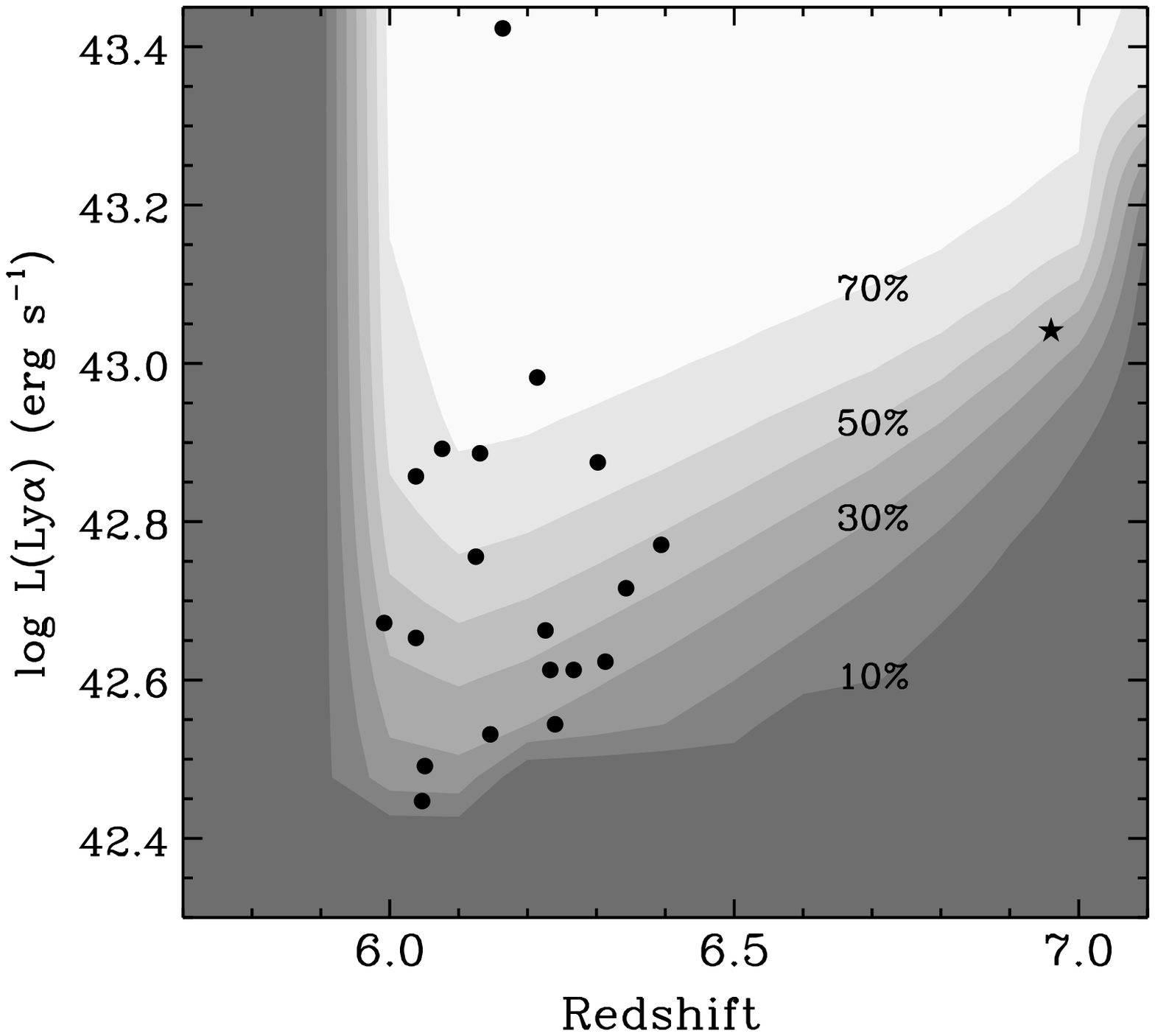}
\caption{LAE selection function as a function of $L$(\lya) and $z$.
The contours are selection probabilities from 10\% to 70\% with an
interval of 10\%. The filled circles are the locations of the 19 $z>6$
galaxies in our sample. The filled star is the $z=6.96$ LAE \citep{iye06}.}
\end{figure}

\begin{figure}
\epsscale{0.7}
\plotone{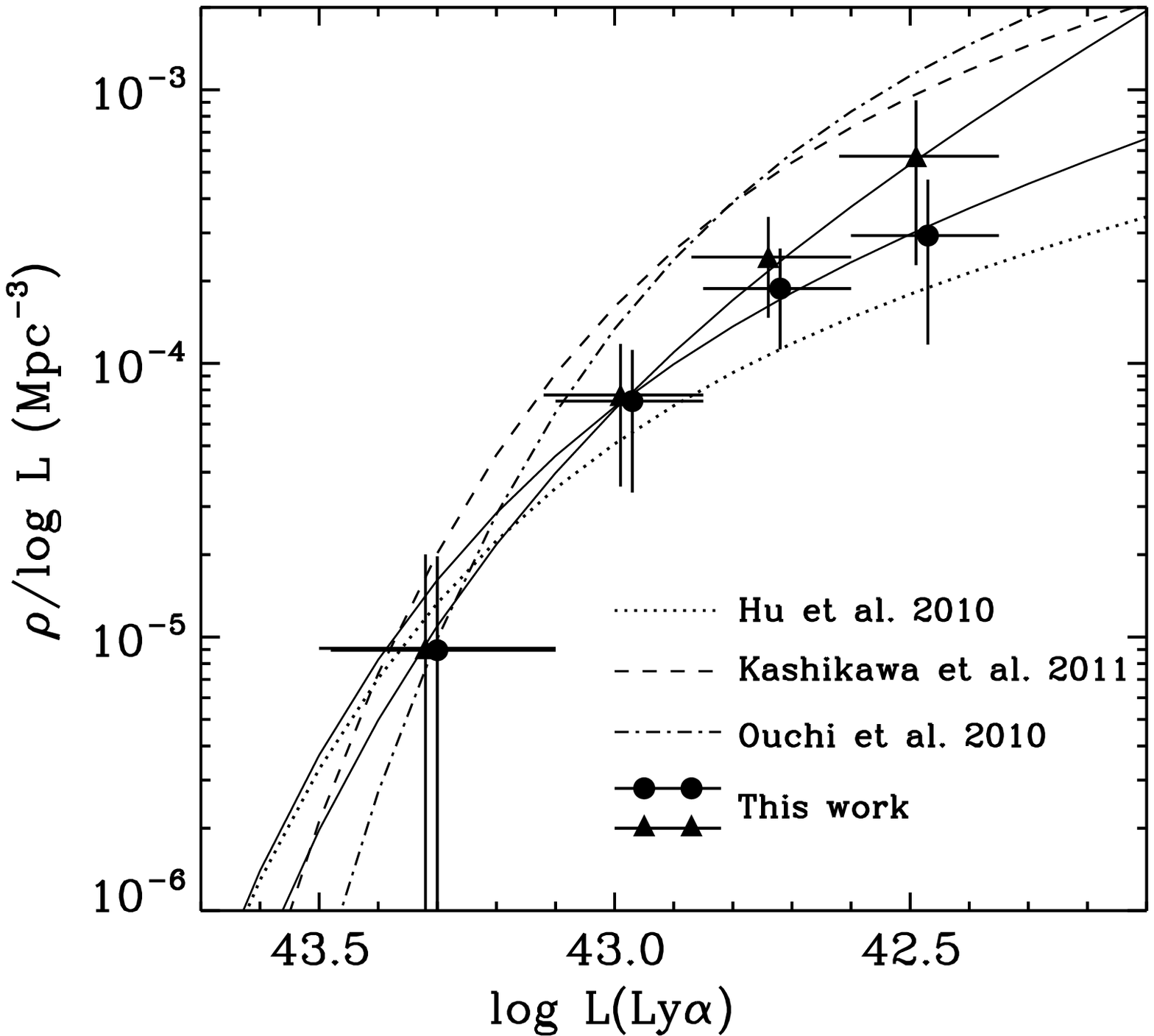}
\caption{\lya\ LF at high redshift. The filled circles and triangles are the 
spatial densities of our sample at $z\sim6.2$ for two different \lya\ EW 
distributions (see Sections 4.3 and 4.4). For the purpose of clarity, the 
triangles have been shifted 0.02 mag along the horizontal axis. The solid 
curves are the best fits to a Schechter function. The dotted, dashed, and
dash-dotted lines represent the \lya\ LFs of LAEs at $z\sim6.5$ from 
\citet{hu10}, \citet{kas11}, and \citet{ouc10}, respectively.}
\end{figure}

\end{document}